\documentclass[twocolumn,twocolappendix]{aastex701}

\usepackage{amsmath}
\usepackage{bm}

\usepackage{breqn}
\definecolor{Orchid}{RGB}{218,112,214}

\def\E{{\cal E}}

\graphicspath{{./}{figures/}}

\newcommand{\hp}[1]{\textcolor{blue}{#1}}

\renewcommand*{\thefootnote}{\fnsymbol{footnote}}

\begin{document}

\interfootnotelinepenalty=10000

\title{Impact of magnetic field-driven anisotropies on the equation of state probed in neutron star mergers}

\author[0000-0002-0491-1210]{Elias R. Most}
\affiliation{TAPIR, Mailcode 350-17, California Institute of Technology, Pasadena, CA 91125, USA}
\affiliation{Walter Burke Institute for Theoretical Physics, California Institute of Technology, Pasadena, CA 91125, USA}
\email{emost@caltech.edu}

\author[0000-0002-6703-418X]{Jeffrey Peterson}
\affiliation{Department of Physics, Kent State University, Kent, OH 44243, USA}
\email{jpeter46@kent.edu}

\author[0000-0002-8763-5660]{Luigi Scurto}
\affiliation{CFisUC, Department of Physics, University of Coimbra, 3004-516 Coimbra, Portugal}
\email{luigiscurto.pr@gmail.com}

\author[0000-0001-7247-1950]{Helena Pais}
\affiliation{CFisUC, Department of Physics, University of Coimbra, 3004-516 Coimbra, Portugal}
\email{hpais@uc.pt}

\author[0000-0001-5578-2626]{Veronica Dexheimer}
\affiliation{Department of Physics, Kent State University, Kent, OH 44243, USA}
\email{vdexheim@kent.edu}

\begin{abstract}
Binary neutron star mergers can produce extreme magnetic fields, some of which can lead to strong magnetar-like remnants. While strong magnetic fields have been shown to affect the dynamics of outflows and angular momentum transport in the remnant, they can also crucially alter the properties of nuclear matter probed in the merger. In this work, we provide a first assessment of the latter, determining the strength of the pressure anisotropy caused by Landau level quantization and the anomalous magnetic moment. To this end, we perform the first numerical relativity simulation with a magnetic polarization tensor and a magnetic-field-dependent equation of state using a new algorithm we present here, which also incorporates a mean-field dynamo model to control the magnetic field strength present in the merger remnant. Our results show that -- in the most optimistic case --  corrections to the anisotropy can be in excess of $10\%$, and are potentially largest in the outer layers of the remnant. This work paves the way for a systematic investigation of these effects. 
\end{abstract}

\keywords{Neutron stars (1108), General relativity(641), Nuclear astrophysics (1129), Nuclear physics (2077)}

\section{Introduction}
From formation \citep{2013ApJ...764..136B}, through early life \citep{Maeder:2003ym,Igoshev:2025alc} and supernova explosions \citep{Wheeler:2001me,Raynaud:2020ist,Barrere:2022gwv}, to different stages of neutron star lives \citep{Bejger:2016emu,Potekhin:2017ufy,Bransgrove:2017jzs,Lander:2020bou, Suvorov:2025ffh,Jiang:2025ijp}, magnetic fields affect many aspects of stellar evolution. Magnetic fields of the order of $10^{16}\,\rm G$ 
\footnote{To convert from Gauss (G) to Gaussian natural units, where the $\sqrt{4\pi}$ appears in the energy-momentum tensor, one can use $1$ MeV$^2$ $= 1.44 \times 10^{13}$ G. To convert to Lorentz-Heaviside natural units, where the $\sqrt{4\pi}$ does not appear in the energy-momentum tensor, one can use $1$ MeV$^2 = 5.11 \times 10^{13}$ G. Alternatively, $m_\pi^2/e \sim 3\times10^{18}$ G.}
have been shown to noticeably deform fully evolved (beyond the proto-neutron star stage) neutron stars \citep{Gomes:2019paw}.
Nevertheless, magnetic fields need to be of the order of $10^{18}$ G to modify the thermodynamical relations (or equation of state, EoS) in the core of neutron stars and proto-neutron stars \cite{Chatterjee:2014qsa,Franzon:2015sya, Strickland:2012vu}. In the crust, this threshold value is lower, $B\gtrsim10^{15}$ G, which has been shown to affect the sub-saturation EoS, originating an extension of the inner crust 
\citep{Fang:2016kcm,Fang:2017zsb,Fang:2017zyz,Avancini:2018jfu,Pais:2021apl,Wang:2022sxx,Scurto:2022vqm}. 
Furthermore, strong modifications of the neutron star crust could have implications for wave launching \citep{Bransgrove:2020rvc}, crustal oscillations \citep{pons2019magnetic}, and connections to radio transients \citep{Bochenek:2020zxn,CHIMEFRB:2020abu}. 
So far, magnetic fields of up to $10^{15}-10^{16}\,\rm G$ have been identified in isolated neutron stars (magnetars) \citep{Kaspi:2017fwg,Makishima:2014dua,Makishima:2018pmu,Makishima:2021vvv,Makishima:2024lnd} and even larger ones could be achieved in their cores \citep{Ferrer:2010wz}. 

Complementary to observing neutron stars, neutron star mergers  \citep{LIGOScientific:2017vwq,LIGOScientific:2020aai} have the potential to constrain properties for cold \citep{Chatziioannou:2018vzf,Raithel:2018ncd,Annala:2017llu,Most:2018hfd,LIGOScientific:2018cki,Bauswein:2017vtn,Margalit:2017dij,Rezzolla:2017aly,Ruiz:2017due,Shibata:2019ctb,Nathanail:2021tay,Tan:2020ics,Most:2020bba,Fattoyev:2020cws} and hot nuclear matter \citep{Bauswein:2010dn,Perego:2019adq,Figura:2020fkj,Raithel:2021hye,Fields:2023bhs,Raithel:2023gct,Raithel:2023zml,Villa-Ortega:2023cps,Miravet-Tenes:2024vba}, including the effect of neutrinos \citep{Alford:2017rxf,Most:2021zvc,Zappa:2022rpd,Espino:2023dei,Most:2022yhe}.
One aspect that has so far been neglected in neutron-star merger simulations is the feedback of strong magnetic fields on the EoS, which has been addressed in isolated stars in many works \citep{Fraga:2008qn,Orsaria:2010xx,Gorbar:2011ya,Dexheimer:2011pz,Aguirre:2013oxa,Sinha:2013dfa,Chu:2014pja,Haber:2014zba,Aguirre:2014rza,Isayev:2014rch,Chamel:2015fna,Fogaca:2016mnw,Tolos:2016hhl,Dexheimer:2016yqu,Avancini:2018jfu,Lugones:2018qgu,Mishra:2018ogw,Chatterjee:2018ytb,Lo:2020ptj,Ferrer:2020tlz,Backes:2021mdt,Lu:2022khf,Prasad:2022dom,Benvenuto:2023smt,Peterson:2023bmr,Mondal:2023baz,Wang:2024lzt,Mondal:2024eaw,Kawaguchi:2024edu}.
While such effects commonly appear only at magnetic fields above $10^{15}\,\rm G$, making them subdominant for most isolated neutron stars, the merger of two neutron stars has been shown to feature dynamo amplification capable of magnetic fields in excess of this limit within milliseconds after merger \citep{Price:2006fi,Kiuchi:2015sga}. 

Using numerical relativity simulations, these dynamo processes have been investigated extensively using either ab-initio \citep{Kiuchi:2017zzg,Aguilera-Miret:2021fre,Chabanov:2022twz,Aguilera-Miret:2023qih,Kiuchi:2023obe} or sub-grid dynamo approaches \citep{Giacomazzo:2014qba,Palenzuela:2015dqa,Most:2023sft,Most:2023sme} to produce strong field strengths. While the background dynamics of magnetic fields in mergers (including winding and breaking \citep{Shapiro:2000zh}) have been extensively investigated \citep{Anderson:2008zp,Giacomazzo:2010bx,Giacomazzo:2013uua,Kiuchi:2014hja,Palenzuela:2015dqa,Endrizzi:2016kkf,Kawamura:2016nmk,Ciolfi:2017uak,Combi:2022nhg,Bamber:2024qzi,Gutierrez:2025gkx}, only recently has it been realized that strong magnetic field in the outer layers of the neutron star can aid the launching of magnetically driven winds, jets {and flares}  \citep{Most:2023sft,Combi:2023yav,Kiuchi:2023obe,Most:2023sme,Jiang:2025ijp,Musolino:2024sju,Bamber:2024kfb}, potentially affecting the electromagnetic afterglow \citep{Metzger:2018qfl,Mosta:2020hlh,Curtis:2023zfo,Combi:2023yav}.
Magnetic fields might also impact angular momentum transport inside the remnant \citep{Margalit:2022rde,Tsokaros:2024wgb,Bamber:2024qzi,Reboul-Salze:2024jst}.
This necessarily requires the outer layers of the magnetar merger remnant to reach field strengths in excess of $10^{16}\, \rm G$ \citep{Kiuchi:2023obe,Most:2023sme}.

In this work we provide a first investigation of the impact magnetic fields could have on the EoS in neutron star mergers. To decrease the model dependency of our work, we study different EoS combinations to model entire neutron stars, from outer crusts to inner cores, including magnetic-field effects, both from Landau quantization and anomalous magnetic moment (AMM) corrections.
Using these equations of state, we then perform fully general-relativistic neutron star mergers simulations with a magnetic polarization tensor and a mean-field dynamo prescription to control the amount of magnetic field amplification produced in the merger. This allows us to systematically vary the field strength in different regions of the merger to disentangle the inherent modifications to the dynamics due to the mere inclusion of strong magnetic fields from the specific feedback on the EoS they may have.

\section{Equation of State Description}\label{sec:eos}

To construct different EoSs, we combine different crust descriptions for the lower density part (containing nuclei, referred to as the crust) and different descriptions for the higher density part (with bulk matter, referred to as the core), to obtain two complete description of neutron stars including the effects of strong magnetic fields: Landau quantization and anomalous magnetic moment (AMM). Note that not many EoSs are available that include magnetic field effects, especially including the AMM. The details on the EoSs we use are given in Appendix~\ref{EOS}, but we summarize them here, and discuss how they generate magnetic-field induced anisotropies in dense matter in the following.

In this study we use two different models: NL3$\omega\rho$ \citep{Horowitz:2000xj,Horowitz:2001ya,Pais:2016xiu} and CMF  \citep{Dexheimer:2008ax, Dexheimer:2011pz}.

The NL3$\omega\rho$ is a Walecka-type of relativistic model \citep{Mueller:1996pm,Serot_1984} that describes nucleons (protons and neutrons) interacting through a mean-field of three mesons: the scalar isoscalar $\sigma$, the vector isoscalar $\omega$ and the vector isovector $\rho$. This model is based on the NL3 interaction \citep{Lalazissis:1996rd}, where  
the additional $\omega\rho$ term was added to ensure a better agreement with astrophysical data \citep{Dexheimer:2018dhb}, by modeling the density dependence of the symmetry energy \citep{Pais:2016xiu}. 
For the crust, we combine an inner crust calculated from a compressible liquid drop model (see e.g. \cite{Lattimer:1985zf,Lattimer:1991nc,Baym:1971ax,Bao:2014lqa,Pais:2015xoa}) under strong magnetic fields \citep{Scurto:2022vqm}
with a smooth transition to a SLy4 outer crust \citep{Douchin:2001sv}.

We also use a hadronic version of the Chiral Mean Field (CMF) model. It is a relativistic model that describes the baryon octet (nucleons and hyperons) also interacting through a mean field of mesons, but now reproducing chiral symmetry restoration \citep{Dexheimer:2008ax, Dexheimer:2011pz}. We combine it with the SLy4 EoS for the outer crust \citep{Douchin:2001sv}.

\begin{figure}[t!]
    \centering
    \includegraphics[width=\linewidth]{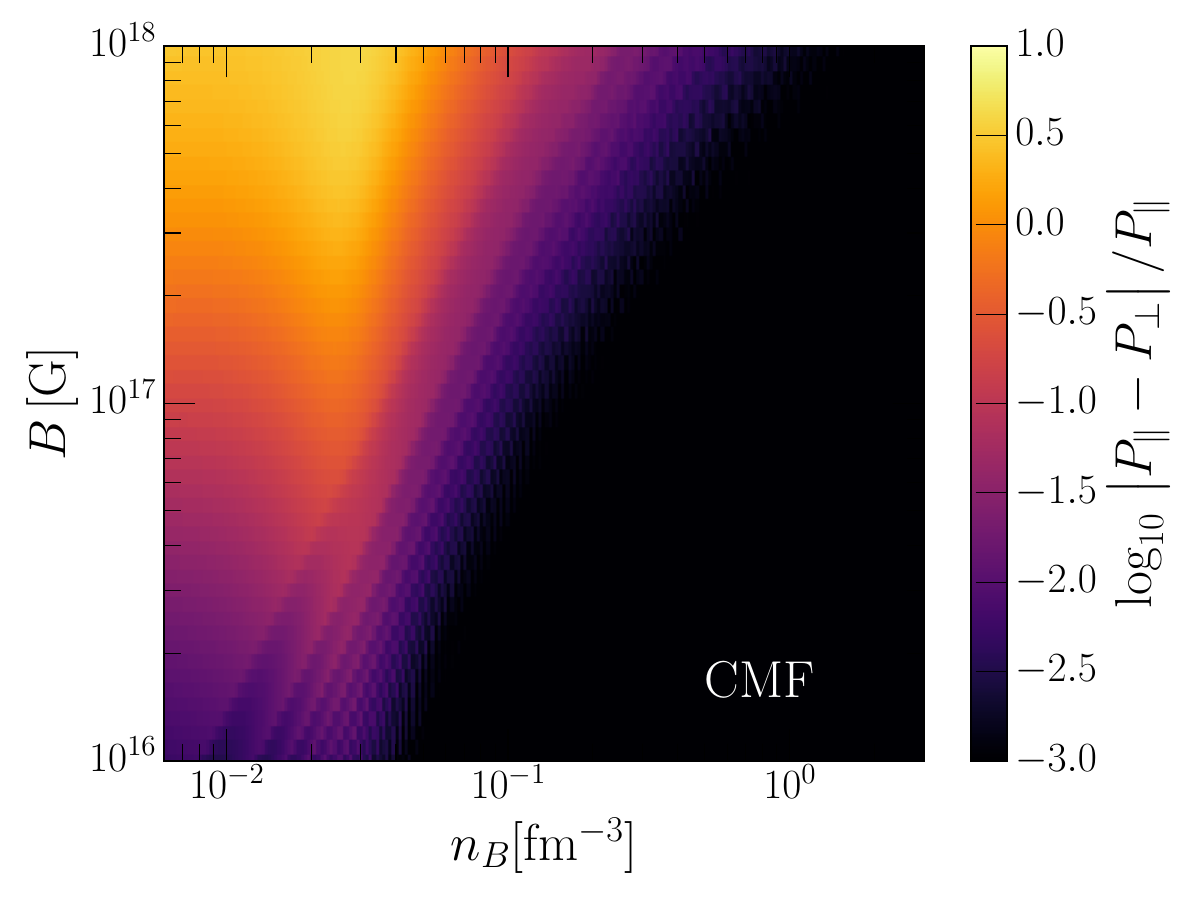}\\
    \includegraphics[width=\linewidth]{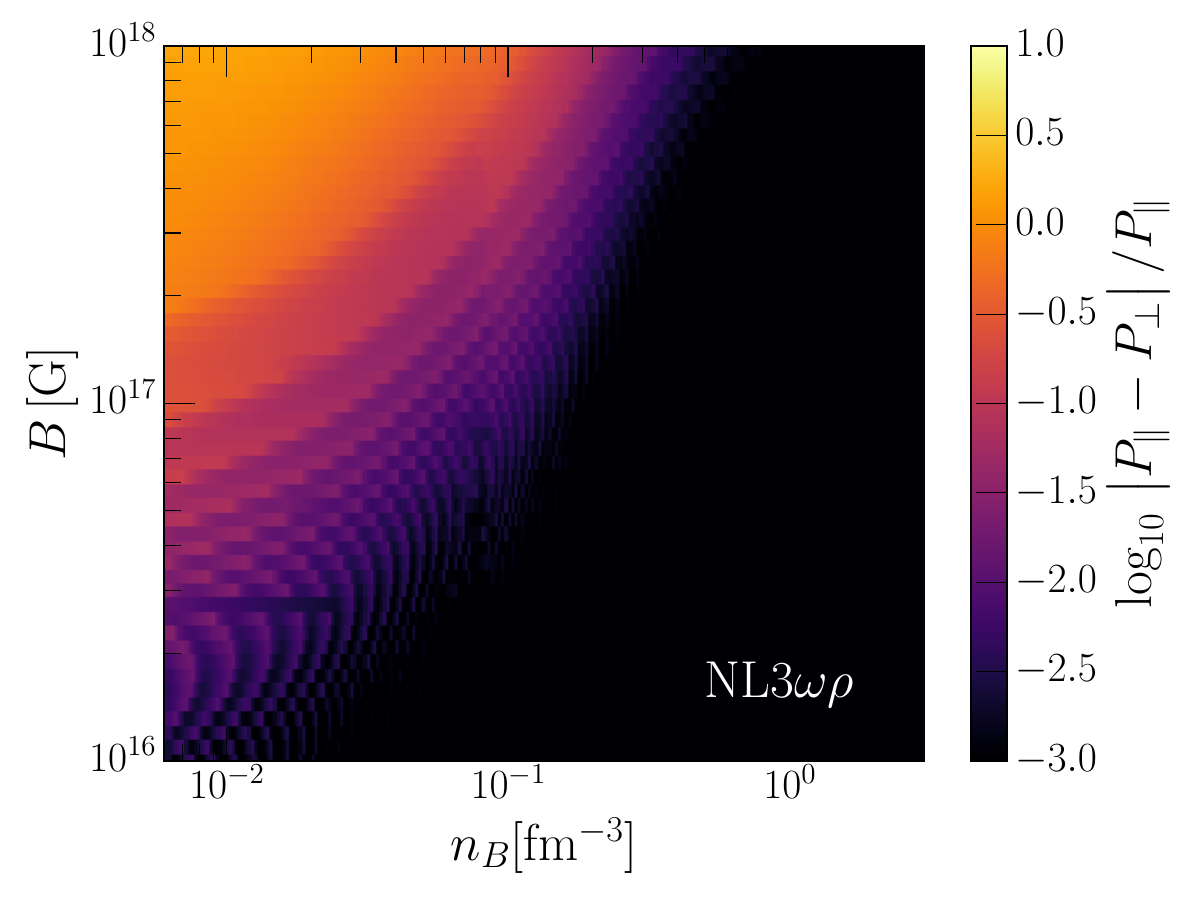}\\
    \caption{{{Normalized} pressure anisotropy, \hp{($P_\parallel - P_\perp)/P_\parallel$}, due to the magnetic field as a function of magnetic-field strength, $B$, and baryon number density, $n_B$, for the CMF model (top panel) and the NL3$\omega\rho$ model (bottom panel). The parallel pressure component, $P_\parallel$, is the thermodynamic pressure, $P_\perp$ is the perpendicular pressure.}}
    \label{fig:eos_overview}
\end{figure}

To highlight how the magnetic field affects differently each core EoS, we show in Fig.~\ref{fig:eos_overview} two-dimensional contours for the pressure anisotropy due to the magnetic field as a function of magnetic-field strength $B$ and baryon (number) density $n_B$. 
We do not extend the figure to very low densities because the outer crust we use does not support a pressure anisotropy.
The pressure anisotropy is defined as the difference between the pressure in the local direction of the magnetic field (parallel pressure) and in the direction perpendicular to it (perpendicular pressure), normalized by the parallel pressure , i.e., $(P_\parallel-P_\perp)/P_\parallel$, where the parallel pressure corresponds to the thermodynamic pressure. In the low magnetic field limit, the anisotropy goes to $0$ (equal pressures, shown in black), while in the extremely high magnetic field case, it goes to $1$ and above (zero perpendicular pressure, shown in orange and negative pressure anisotropy, shown in yellow). The figure shows that the stronger the field, the higher the density it can modify the EoS, with stronger effects overall for the CMF model, which contains hyperons. The color oscillations are caused by the Van Alphen oscillations due to the discrete nature of the Landau levels \citep{DHVA}.

\section{General-relativistic magnetohydrodynamics with magnetic  polarization}

In this study, we solve the equations of general-relativistic magnetohydrodynamics
for use with the magnetic field dependent EoSs described in Sec.
\ref{sec:eos}. Most importantly, we also provide a simple way for including magnetic 
polarization into the equations we solve
\citep{Chatterjee:2014qsa,Pimentel:2018uwh}.
In the following, we 
largely adopt the language of the 3+1 formulation of general-relativity
(e.g., \citet{Gourgoulhon:2007ue}). Consequently, we decompose the spacetime
metric as,
\begin{align}
{\rm d}s^2 =& g_{\mu \nu} {\rm d}x^\mu {\rm d} x^\nu \nonumber\\
=& \left(-\alpha^2 + \beta_i \beta^i\right) + 2 \beta_i {\rm d} t {\rm d} x^i + \gamma_{ij} {\rm d} x^i {\rm d} x^j\,,
\end{align}
where $\alpha$ is the lapse, $\beta^i$ the shift vector, and $\gamma_{ij}$ the induced three-metric on the hypersurface defined by the normal vector, $n_\mu = (-\alpha,0,0,0)$.
We model the dynamics of matter using general-relativistic magnetohydrodynamics in dynamical spacetimes \citep{Baumgarte:2002vv,Duez:2005sf}. Matter is described in terms of an energy density $e$, pressure $P$ and four-velocity $u^\mu$. Within the 3+1 split, we can define a three-vector velocity $v_i = Wu_i$, such that $W = 1/\sqrt{1-v_i v^i}$ is the Lorentz factor.

The electromagnetic sector is described by the vector potential,
\begin{align}
\mathcal{A}_\mu = \Phi n_\mu + A_\mu\,,
\end{align}
where $A_i$ is the spatial component, and $\Phi$ the gauge potential. This gives rise to the the field strength tensor,
\begin{align}
F_{\mu\nu} = \nabla_\mu A_\nu - \nabla_\nu A_\mu\,,
\end{align}
where $\nabla_\mu$ is the four-dimensional covariant derivative, and its dual
$\,^{\ast}\!F^{\mu\nu} = -\frac{1}{2}\varepsilon^{\mu\nu\kappa\lambda} F_{\mu\nu}$.
We can then define electric and magnetic fields in the normal and co-moving frame, via,
$E^\mu = n_\nu F^{\mu\nu}$, $B^\mu = n_\nu \,^{\ast}\!F^{\mu\nu}$, and $e^\mu = u_\nu F^{\mu\nu}$, $b^\mu = u_\nu \,^{\ast}\!F^{\mu\nu}$, respectively.

We then assume the ideal magnetohydrodynamics approximation, $e^\mu \approx 0$, appropriate for the highly electrically conductive regime probed in neutron star mergers \citep{Harutyunyan:2018mpe}. However, since we want to control dynamo amplification in the merger remnant, we supplement it with a mean-field dynamo term \citep{Most:2023sme},
\begin{align}
e^\mu = \kappa b^\mu\,,
\end{align}
where $\kappa$ is the dynamo coefficient \citep{Bucciantini:2012sm}.
This translates to a normal electric field \citep{Most:2023sme},
\begin{align}
E^i =& - \varepsilon^{ijk}v_j B_k + \kappa \left[ \left(1- v^2\right) B^i  +  \left(v_l B^l\right) v^i \right] + \mathcal{O}\left(\kappa^2\right)\,.
\label{eqn:Efield_scalar}
\end{align}
Overall, this leads to a final evolution equation \citep{Most:2023sme},
\begin{align}
\partial_t A_i &= \frac{\alpha}{u^0} \varepsilon_{ijk}u^j {B}^k - \alpha \kappa\left[ \left(1-v^2\right) {B}_i + \left(v_l B^l\right) v_i \right] \nonumber \\
& - \partial_i \left(\alpha \Phi - \beta^j A_j \right)\,,
\label{eqn:A_evol}
\end{align}
where the magnetic field follows from $B^i = \varepsilon^{ijk} \partial_j A_k$, and 
$\varepsilon^{ijk}$ is the Levi-Civita tensor on the hypersurface.

The energy momentum tensor of the resulting dynamo system  is identical to ideal magnetohydrodynamics up to the order we consider \citep{Most:2023sme},
\begin{align}
T^{\mu\nu}_{\rm ideal} =& \left(\varepsilon+P + b^2\right) u^\mu u^\nu + \left(P + \frac{1}{2} b^2\right) g^{\mu\nu} - b^\mu b^\nu \nonumber \\
&+ \mathcal{O}\left(\kappa^2\right)\,,
\end{align}
{where $\varepsilon$ and $P$ {(or $P_\parallel$)} in our simulations comes from Eqs.~\eqref{eq:Fermisums1} and \eqref{eq:Fermisums2} for CMF and Eqs.~\eqref{enerCrust}, \eqref{Eq_energy1} and \eqref{Ppar} for NL3$\omega\rho$.}

Additionally, it has been shown that the presence of strong magnetic fields introduces
an anisotropy in the pressure. This can be captured by tracking the magnetic polarization of the material in terms of a polarization vector $m^\mu$ \citep{Chatterjee:2014qsa}.
The full energy momentum tensor of a polarizable magnetohydrodynamical fluid is given as 
\begin{align}
T^{\mu\nu}= T^{\mu\nu}_{\rm ideal} + T^{\mu\nu}_{\rm mag}\,,
\end{align}
where we now have a correction \citep{Chatterjee:2014qsa,Pimentel:2018uwh},
\begin{align}
T^{\mu\nu}_{\rm mag} = \frac{1}{2}\left[m^\mu b^\nu + m^\nu b^\mu\right]  - \left[ u^\mu u^\nu + g^{\mu\nu}\right] b^\alpha m_\alpha  \,.
\end{align}
{giving rise to a perpendicular pressure ($P_\perp$), which for our scenario comes from Eq.~\eqref{eq:Fermisums3} for CMF and Eq.~\eqref{pperpH} for NL3$\omega\rho$.}
We now assume that the polarization aligns with co-moving magnetic field, i.e.,
\begin{align}
m^\mu = \mu b^\mu\,,
\end{align}
where $\mu = \left(P_\perp - P_\parallel\right)/b^2$ is the magnetic susceptibility.

The non-ideal correction then takes a Braginskii-like form \citep{Chandra:2015iza,Most:2021rhr}
\begin{align}
    T^{\mu\nu}_{\rm mag} = -\frac{2}{3} \mu b^2 \Delta^{\mu\nu} + \mu \left(b^\mu b^\nu - \frac{1}{3} \Delta^{\mu\nu} b^2\right)\,,
\end{align}
corresponding to a bulk pressure and anisotropic pressure correction. However, different from Braginskii theory the anisotropy is not evolved, but provided by the EoS. This allows us to apply a fluid-frame transformation to recast the equation as closely as possible to ideal MHD.
This results in the effective system,
\begin{align}
T^{\mu\nu} = \left(\tilde{e}+\tilde{P} + \tilde{b}^2\right) u^\mu u^\nu + \left(\tilde{P} + \frac{1}{2} \tilde{b}^2\right) g^{\mu\nu} - \tilde{b}^\mu \tilde{b}^\nu\,,
\end{align}
where 
\begin{align}
\tilde{b}^\mu &= \sqrt{1-\mu}\, b^\mu\,,\\
\tilde{e} &= e - \frac{\mu}{2\left(1-\mu\right)} \tilde{b}^2 = e - \frac{P_\perp - P_\parallel}{2}\,, \label{eqn:etilde}\\
\tilde{P} &= P_\parallel - \frac{\mu}{2\left(1-\mu\right)} \tilde{b}^2 = \frac{3}{2} P_\parallel - \frac{1}{2} P_\perp \label{eqn:btilde}\,.
\end{align}
The evolution of the system we solve is then governed by
\begin{align}
\nabla_\mu \left(\rho u^\mu\right) &=0\,, \\
\nabla_\mu T^{\mu\nu} &=0\,,\\
\nabla_\mu\!^{\ast}\!F^{\mu\nu} &=0\,,
\end{align}
where $\rho= m_B n_B$ is the baryon rest-mass density, $m_B$ is the baryon mass.

\begin{figure*}[t!]
    \centering
    \includegraphics[width=0.9\linewidth]{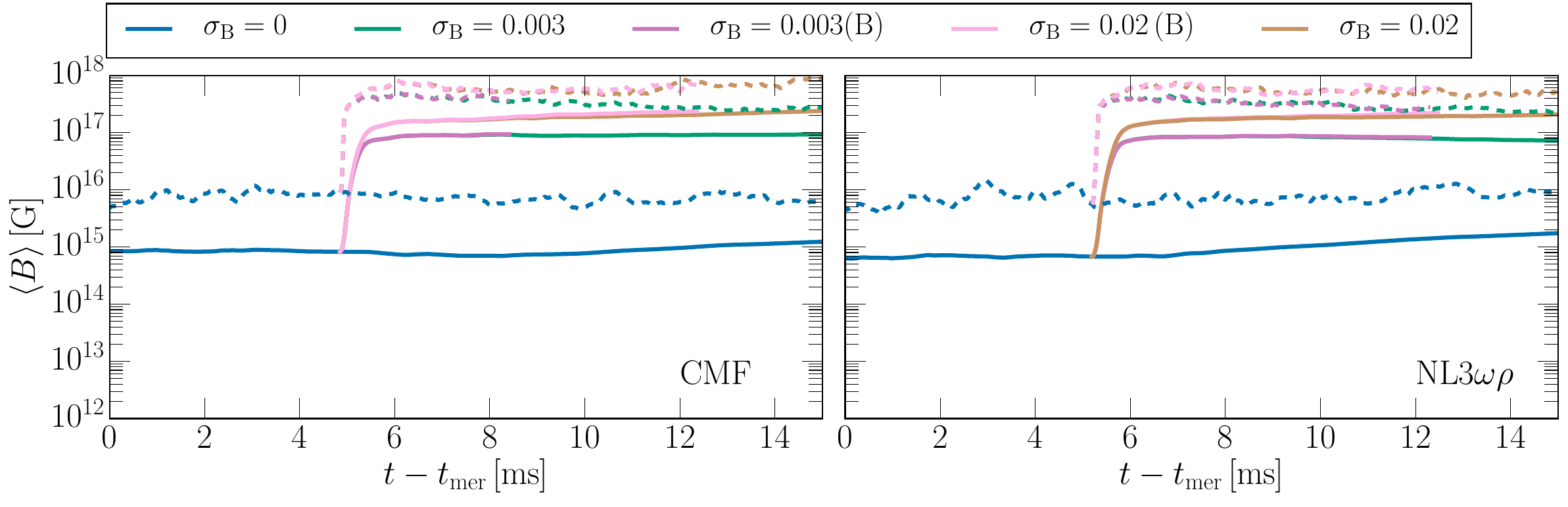}
    \caption{Evolution of the magnetic field strength, $B$, during and after merger. Shown are mass-weighted averages (solid lines) and maximum values (dashed lines) for both equations of state, CMF and NL3$\omega\rho$, where `B' indicates feedback of the magnetic field onto the equation of state. $\sigma_B$ indicates the magnetization for saturation of the mean-field dynamo. Times, $t$, are stated relative to the time of merger, $t_{\rm mer}$.}
    \label{fig:b_evol}
\end{figure*}

We can gain some insights into the meaning of this pressure correction by using the language of out-of-equilibrium hydrodynamics \cite{Romatschke:2009im}.
{We can see that this transformation is akin to choosing a generalized hydrodynamic frame, which features out-of-equilibrium energy and pressure corrections, $\tilde{P} = P_\parallel + \Pi$ \citep{Noronha:2021syv,Rocha:2021lze}.}
In particular, we may use that the pressure anisotropy, $\Pi$, takes the form of an effective bulk pressure correction,
\begin{align}
    \Pi = \frac{1}{2}\left(P_\parallel - P_\perp\right)\,,
\end{align}
which satisfies
\begin{align}
    u^\mu \nabla_\mu \Pi = &- \left(\rho \left.\frac{\partial \Pi}{\partial \rho}\right|_{b^2} + 2 b^2 \left.\frac{\partial \Pi}{\partial b^2}\right|_{\rho} \right) \nabla_\mu u^\mu \nonumber \\ 
    &+ 2 \left.\frac{\partial \Pi}{\partial b^2}\right|_{\rho} b_\mu b_\nu \sigma^{\mu\nu}\,,
\end{align}
where $\sigma_{\mu\nu}$ is the shear tensor. Different from, e.g., neutrino-driven bulk viscosity \citep{Most:2021zvc,Most:2022yhe,Gavassino:2023xkt}, the pressure anisotropy is a non-dissipative (damping) effect.
As we will see, the anisotropy largely builds up due to increase in magnetic field strength after the initial collision and core bounces, where compression is largest \citep{Nedora:2021eoj}.
As such, the anisotropy is purely sourced by shear-driven (dynamo) amplification of the magnetic field
\begin{align}\label{eqn:Pi_evol}
    u^\mu \nabla_\mu \Pi \approx  2 \left.\frac{\partial \Pi}{\partial b^2}\right|_{\rho} b_\mu b_\nu \sigma^{\mu\nu}\,,
\end{align}
where the above holds approximately in the conditions of the post-merger.

\begin{figure*}[t!]
    \centering
    \includegraphics[width=0.8\linewidth]{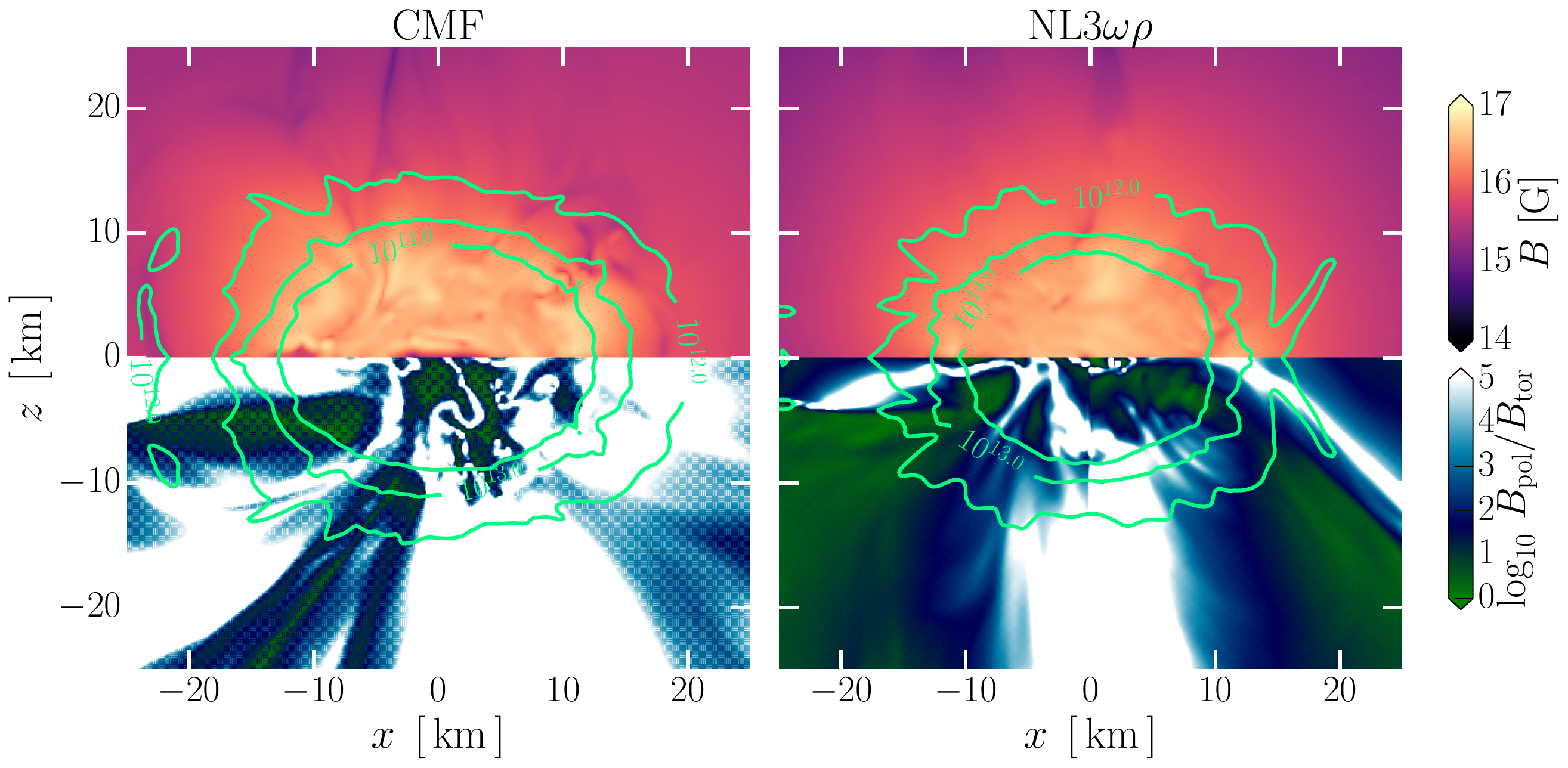}
    \caption{Hypermassive neutron star using the CMF and NL3$\omega\rho$ equations of state about $12\, \rm ms$ after merger. Shown is the absolute magnetic field strength, $B$, (top panels) and the ratio of poloidal, $B_{\rm poloidal}$, and toroidal, $B_{\rm toroidal}$, magnetic field (bottom panels). Green contour lines indicate level surfaces of the baryon rest-mass density $\rho_B$ in $\rm g/cm^3$.}
    \label{fig:hmns}
\end{figure*}

While magneto-turbulence is prevalent throughout the entire remnant after merger \citep{Aguilera-Miret:2021fre}, the saturation of the dynamo requiring near-equipartition values for the anisotropy to become dynamically important, $\Pi \gtrsim P_\parallel$.
This makes the outer layers of the merger remnant, where a turbulent $\alpha\Omega-$dynamo operates \citep{Kiuchi:2023obe}, a most promising site, as the field strength there easily reach equipartition \citep{Most:2023sme}, aiding breakout of the field from the surface of the star \citep{Most:2023sft,Combi:2023yav,Most:2023sme,Musolino:2024sju}.

\section{Numerical implementation}

We numerically solve the equations of ideal GRMHD with polarizable matter
in dynamical spacetimes using the \texttt{Frankfurt/IllinoisGRMHD (FIL)}
code \citep{Most:2019kfe}, and its recently developed dynamo infrastructure
\citep{Most:2023sme}. In detail, we solve the Einstein field equations in
the Z4c formulation \citep{Bernuzzi:2009ex,Hilditch:2012fp} in moving
puncture gauge \citep{Alcubierre:2002kk} using a fourth-order accurate
finite-difference discretization \citep{Zlochower:2005bj}. The GRMHD
equations are solved using a fourth-order accurate variant of the
conservative finite-difference scheme ECHO scheme \citep{DelZanna:2007pk},
including a robust primitive inversion scheme for high magnetization
\citep{Kastaun:2020uxr}, which we have modified for use with the magnetic field dependent EoSs, see Appendix~\ref{app:inversion}. The equations of state are tabulated.
Additional details on the code and setup can be found in \citealt{Most:2019kfe,Most:2021ktk,Raithel:2022orm}.

We consider different magnetic-field-dependent equations of state at zero-temperature, CMF and NL3$\omega\rho$, as described in Sec. \ref{sec:eos}. In order to compare the effect of magnetic field dependence, we perform two sets of simulations, one with magnetic field feedback included, the other one fixed to its zero magnetic field slice. For these sets of simulations the impact of pressure anisotropy is then only estimated in post-processing. We also use two dynamo amplification parameters, such that the dynamo saturates at a given level of target magnetization $\sigma_B = b^2/\rho = \{0.003; 0.02\}$ \citep{Most:2023sft}.
The initial data for our simulations is computed using the $\texttt{FUKA/Kadath}$ library \citep{Papenfort:2021hod,Grandclement:2009ju}. Specifically we adopt non-spinning equal mass configurations with a total mass of $2.6\, \rm M_\odot$.
We further adopt a simulation domain spanning $[-2,048:2,048]\,\rm km$ over $8$ refinement levels, with a finest resolution of $250\,\rm m$.

\section{Results}

\begin{figure*}[t!]
    \centering
    \includegraphics[width=\linewidth]{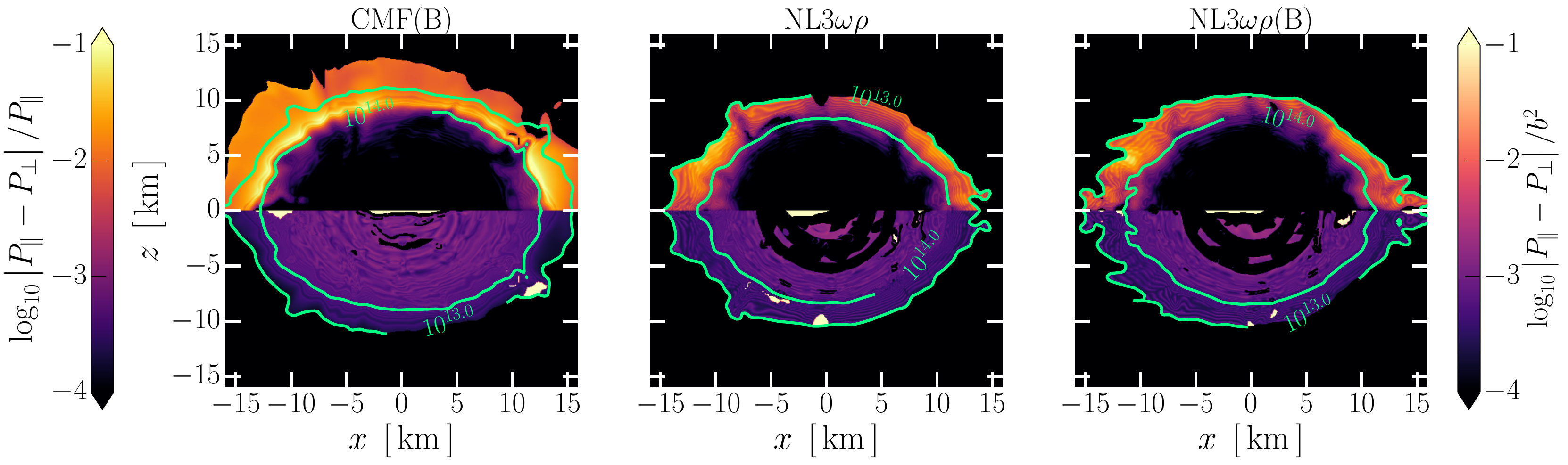}
    \caption{Normalized pressure anisotropy, $(P_\parallel - P_\perp)/P_\parallel$, (top panels) and {magnetic susceptibility}, $\mu = (P_\parallel - P_\perp)/b^2$, (bottom panels) shown at $8\,\rm ms$ after merger for the CMF and NL3$\omega\rho$ equation of state. $b^2$ denotes the local magnetic field energy in the comoving frame. {We show results for the equation of state (EOS) with magnetic field dependence ('B'), and post-processed when simulating only with the $B=0$ EOS}.}
    \label{fig:deltap_hmns}
\end{figure*}

In this work, we want provide a first assessment of the impact of magnetic-field driven pressure anisotropies in binary neutron star mergers.
Before presenting this main result of our study (Sec. \ref{sec:aniso}), we provide a brief description of the merger dynamics, which has largely been investigated in previous work (see, e.g., \citet{Kiuchi2025} for a recent review).

During the merger of two neutron stars, a shear layer develops at the contact interface \citep{Price:2006fi,Baiotti:2008ra}. It has been demonstrated that this shear layer, together with the onset of turbulence throughout the merger remnant leads to an amplification of initially weak magnetic fields to values of $10^{17}\, \rm G$ in the central regions of the merger remnant \citep{Kiuchi:2015sga,Palenzuela:2022kqk,Aguilera-Miret:2023qih}.
After the merger, a remnant develops which has a nearly uniformly rotating inner core, a rapidly rotating outer layer with radially increasing angular velocity, and a Keplerian envelope at crustal densities \citep{Hanauske:2016gia}. This latter part has been shown to be unstable to the magnetorotational instability, developing an effective $\alpha\Omega$-dynamo \citep{Kiuchi:2023obe}, that can drive the outer layers to an equipartition state between the magnetic field energy and the fluid, leading to an effective breakout of the field from the star \citep{Most:2023sft,Combi:2023yav}. 
As a consequence, those regions are in a regime where magnetic field corrections to the EoS could become important, although we caution that finite-temperature effects (which we do not self-consistently capture) may alter this picture \citep{Strickland:2012vu}.

Since our simulations are not able to fully capture the dynamo amplification self-consistently, we utilize a mean-field dynamo model \citep{Most:2023sme} to mimic the amplification process outlined above. 
Starting from the post-merger remnant around $t=5\,\rm ms$ after merger (Fig. \ref{fig:b_evol}), we activate the mean-field dynamo term, and drive the system into a state of {constant magnetization $\sigma_B = b^2/\rho = \left\{0.003; 0.02\right\}$ (see, e.g., \citep{Most:2023sft,Most:2023sme}}, which we vary to capture different magnetic field amplification strengths. {Note that subsequent self-consistent evolution after the dynamo term is deactivate will further enhance the magnetic field strength, making it overall more similar for both configurations.}
We generally consider two cases: In the first case, we consider dynamical evolutions using the $B=0$ part of the EoS, and then estimate in post-processing how large the pressure anisotropy would be. In the second part, we perform fully backreacted simulations using the magnetic field dependent EoS and the polarization tensor. 
The subsequent evolution beyond this initial saturation proceeds self-consistently in ideal GRMHD.
We show this evolution in Fig. \ref{fig:b_evol}, where we depict the mass weighted average $\left<B\right>$ of the magnetic field strength. As soon as we activate the dynamo, the magnetic field inside the remnant is quickly amplified. Since we choose a uniform $\sigma_B$ amplification \citep{Most:2023sft}, rather than a sophisticated model for the $\alpha\Omega-$dynamo \citep{Most:2023sme}, the field strength is amplified throughout the remnant and quickly saturates. 

The overall evolution is relatively insensitive to the EoS, barring intrinsic differences due to changes in stellar structure. In general, we find that without the addition of a dynamo term, our simulations saturate at field strengths between $10^{15}-10^{16}\,\rm G$. If a mean-field dynamo term is included, we easily reach average magnetic field strengths of $10^{16}-10^{17}\, \rm G$ inside the remnant. The maximum field strength probed, depends on the exact dynamo saturation value and is in excess of $10^{17}\,\rm G$, which is sufficient to trigger pressure anisotropies as discussed in Sec. \ref{sec:eos}. Since the pressure anisotropy depends critically on the magnetic field topology, Fig. \ref{fig:hmns} shows the relative ratio of poloidal, $B_{\rm poloidal}$ and toroidal, $B_{\rm toroidal}$, field inside the magnetar remnant. We can see that the remnant features a mix of toroidal and poloidal field, with strong poloidal fields present at the surface of the neutron star, leading to breakout of the field from the remnant \citep{Most:2023sft,Combi:2023yav,Musolino:2024sju}.

\subsection{Pressure anisotropy}\label{sec:aniso}

\begin{figure*}[t!]
    \centering
    \includegraphics[width=0.9\linewidth]{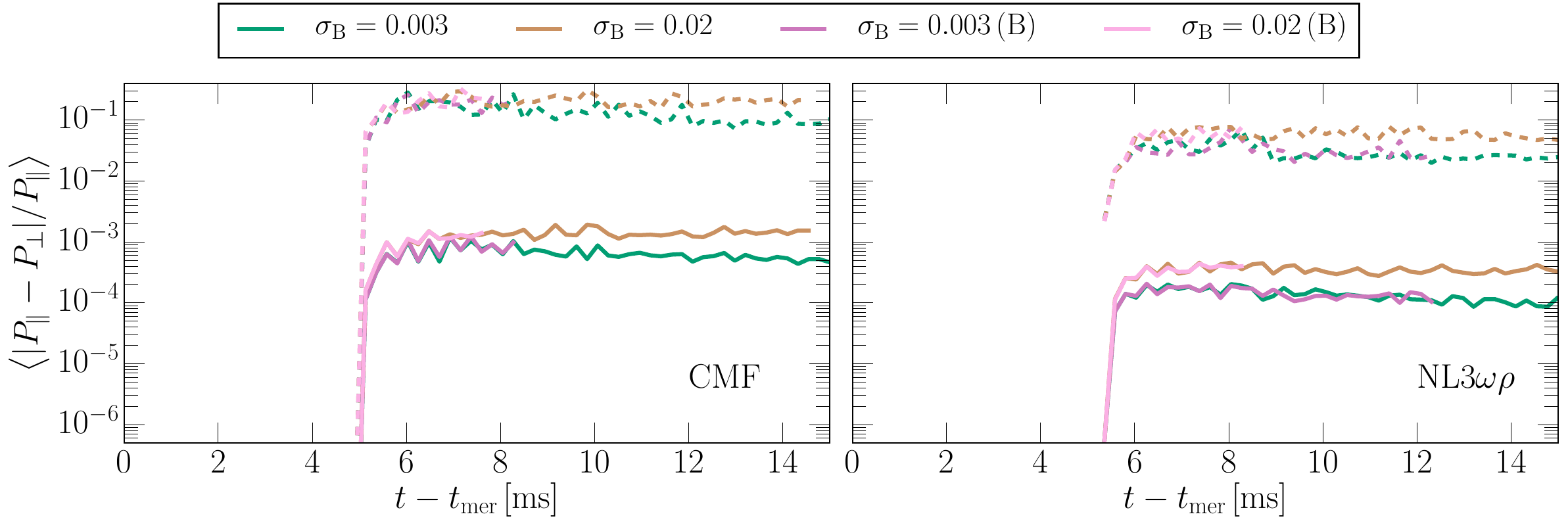}
    \caption{Magnetic pressure anisotropy inside the magnetar remnant. Shown are mass-weighted averages (solid lines) and maximum values (dashed lines) for both equations of state, CMF and NL3$\omega\rho$. The maximum values are always located inside the crust region, whereas the mass-weighted averaged provide a faithful representation inside the core. Different colors denote dynamo saturation levels in terms of the magnetization $\sigma_B$. The letter `B' indicates a simulation with magnetic field feedback on the equation of state included. }
    \label{fig:deltap_evol}
\end{figure*}

Owing to the large magnetic field strength we drive in the hypermassive neutron star, a net Braginskii-like pressure anisotropy, $\left(P_\parallel - P_\perp \right)$ builds up in the system, as can be seen from Eq. \eqref{eqn:Pi_evol}. 
We quantify this pressure anisotropy around $8\,\rm ms$ after merger in Fig. \ref{fig:deltap_hmns}. 
We can see that for sufficient magnetic field amplification, pressure anisotropies of order unity can be probed in the outer layers of the remnant. While they are also present at high density, the magnetic field strengths we probe there are further away from their equipartition values \citep{Most:2023sme}, leading to only small anisotropies of order $\Pi \simeq 10^{-3}-10^{-4} P_\parallel$, {with the spread being determined by the choice of equation of state}. We caution that, overall, this strongly depends on the magnetic field strengths produced in the merger, which can be larger than what we use here \citep{Kiuchi:2017zzg,Aguilera-Miret:2021fre}.

In Fig. \ref{fig:deltap_evol}, we also quantify the temporal evolution of the pressure anisotropy. As demonstrated in Eq. \eqref{eqn:Pi_evol}, the anisotropy does not necessarily decay, except if the field was strongly sheared apart. This makes this effect unlike conceptually similar pressure corrections, such as Urca-driven bulk viscosity \citep{Most:2021zvc,Most:2022yhe}. Accordingly, we find that the average pressure anisotropy initially increases during merger, reaching values of $\Pi \simeq 10^{-3} P_\parallel$ primarily inside the core. Such values are likely too small to affect the subsequent evolution of the system, as they are smaller than thermal \citep{Figura:2020fkj,Raithel:2021hye,Fields:2023bhs} and bulk viscous corrections \citep{Most:2021ktk,Most:2022yhe}. {However, inside the crustal regions of the remnant, anisotropies can reach between $\Pi \simeq \left(10^{-2}-10^{-1}\right) P_\parallel$} (as we have seen in Fig. \ref{fig:deltap_hmns}). Overall, we can see that the evolution of the anisotropy, both maximum values (in the crust) and average values (in the core), are comparable between the two sets of simulations including the backreaction of the field on the EoS or not.
These conclusions seem largely independent of the high-density EoS, {and primarily depend on physics below saturation densities, which for the densities considered for this effect is different for the two systems considered}.
We caution, however, that we neglect finite-temperature corrections to the EoS, which naturally lowers the expected anisotropy. As such our results, should be seen as an upper bound of this effect.

\section{Conclusions}

Magnetic fields in neutron star mergers can be amplified to magnetar-level field strengths \citep{Kiuchi:2017zzg,Aguilera-Miret:2023qih,Kiuchi:2023obe,Most:2023sme}. For those values, the magnetic field can become large enough to affect the properties of dense matter via Landau level quantization, and the anomalous magnetic moment \citep{DHVA}.
We have presented the first assessment of this anisotropy on the post-merger dynamics of a binary neutron star coalescence. In doing so, we have incorporated a polarization tensor into the GRMHD equations using the formulation of \citet{Chatterjee:2014qsa}, which allows us to control the pressure anisotropy present in nuclear matter. We have then used two different descriptions of nuclear matter, \texttt{CMF} and \texttt{NL3}$\omega\rho$, both of which include an effective magnetic field dependence and a crust.
By performing fully general-relativistic simulations of binary neutron star coalescence, we demonstrated that after merger magnetic fields, in particular in the outer regions of the merger remnant become strong enough to reach the threshold for magnetic field anisotropies to {potentially} become dynamically important. We find that locally the anisotropy can reach values of $10\%$, for the strongest magnetic field amplification strength in the post-merger remnant (this number is EoS dependent), which in the relevant regions is largely driven by the $\alpha\Omega$-dynamo \citep{Kiuchi:2023obe,Most:2023sme}. Since these regions are susceptible to eventual breakout of the magnetic field from the star by means of Parker instabilities \citep{Most:2023sft,Combi:2023yav,Kiuchi:2023obe,Musolino:2024sju,Jiang:2025ijp}, the anisotropy may change the dynamics of jet and wind launching from these systems. A detailed analysis of this aspects will be presented in future work.
One aspect not presently addressed is the temperature dependence of the anisotropy \citep{Strickland:2012vu}. Our initial assessment has so far relied on a fixed cold segment of the EoS, that was augmented with a thermal component, which did not couple to the magnetic field. A self-consistent analysis should be carried out in future work to clarify this dependence.

\section{Acknowledgments}

The authors are grateful for discussions with Constan\c ca Provid\^encia and Ira Wasserman.
ERM acknowledges support from NASA's ATP program under grant 80NSSC24K1229
and by the National Science Foundation under grants No. PHY-2309210. The work of LS and HP was partially supported by national funds from FCT (Fundação para a Ciência e a Tecnologia, I.P, Portugal) under projects UIDB/04564/2020 and UIDP/04564/2020, with DOI identifiers 10.54499/UIDB/04564/2020 and 10.54499/UIDP/04564/2020, respectively, and the project 2022.06460.PTDC with the associated DOI identifier 10.54499/2022.06460.PTDC. LS acknowledges the PhD grant 2021.08779.BD (FCT, Portugal). HP acknowledges the grant 2022.03966.CEECIND (FCT, Portugal) with DOI identifier 10.54499/2022.03966.CEECIND/CP1714/CT0004. V.D. acknowledges support from the Department of Energy under grant DE-SC0024700 and from the National Science Foundation under grants MUSES OAC-2103680 and NP3M PHY2116686.
ERM acknowledges the use of Delta at the National Center for Supercomputing Applications (NCSA) through allocation PHY210074 from the Advanced Cyberinfrastructure Coordination Ecosystem: Services \& Support (ACCESS) program, which is supported by National Science Foundation grants \#2138259, \#2138286, \#2138307, \#2137603, and \#2138296. Additional simulations were performed on the NSF Frontera supercomputer under grant AST21006. ERM also acknowledges support through DOE NERSC supercomputer Perlmutter under grant m4575, which uses resources of the National Energy Research Scientific Computing Center, a DOE Office of Science User Facility supported by the Office of Science of the U.S. Department of Energy under Contract No. DE-AC02-05CH11231 using NERSC award NP-ERCAP0028480.

\software{ 
    EinsteinToolkit \citep{Loffler:2011ay},
    Frankfurt/IllinoisGRMHD \citep{Most:2019kfe,Etienne:2015cea}
    FUKA \citep{Papenfort:2021hod},
    Kadath \citep{Grandclement:2009ju},
    kuibit \citep{kuibit},
	  matplotlib \citep{Hunter:2007},
	  numpy \citep{harris2020array},
	  scipy \citep{2020SciPy-NMeth}
}

\appendix
\setcounter{figure}{0}
\renewcommand{\thefigure}{\Alph{section}.\arabic{figure}}

\section{Equations of State}
\label{EOS}

\begin{figure}
    \centering
    \includegraphics[width=\linewidth]{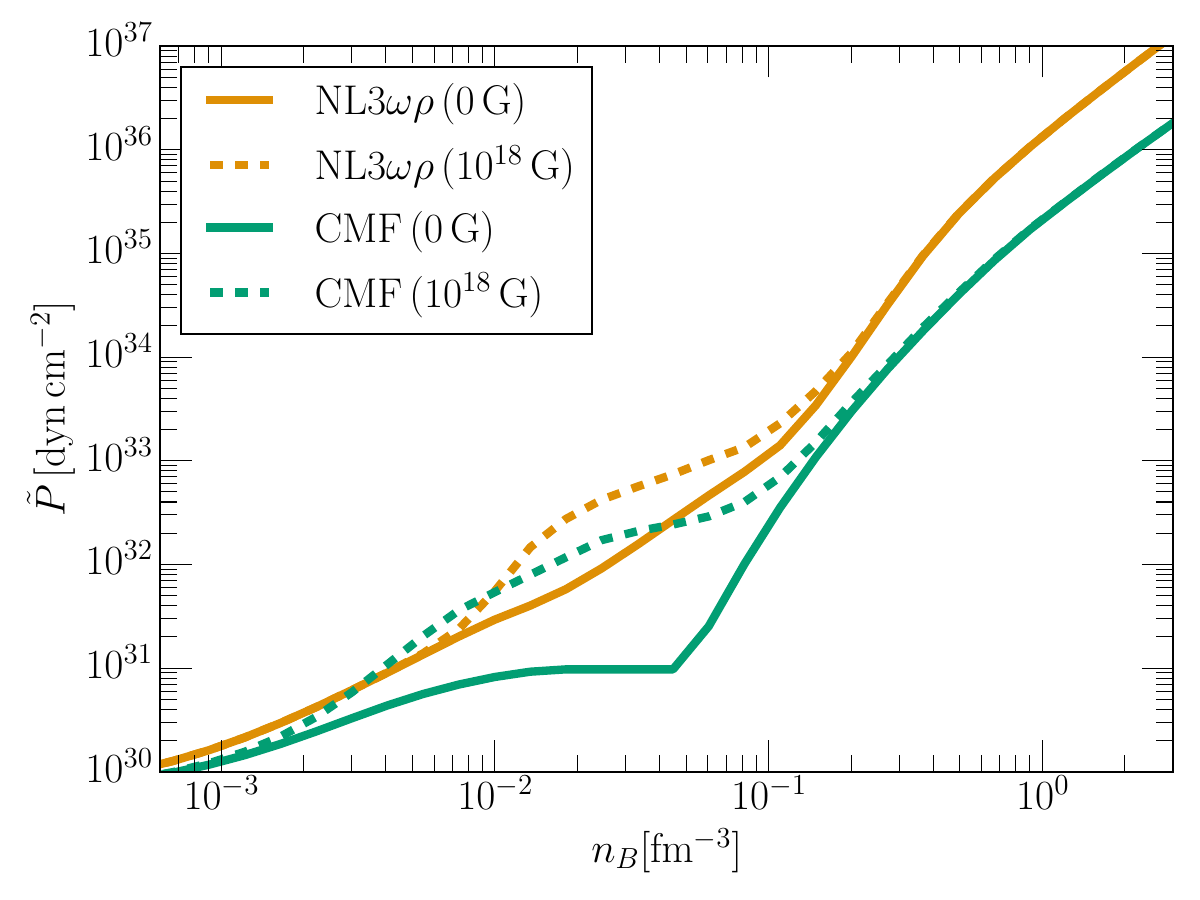}
    \caption{Equations of state used in this work. Shown are the $CMF$, NL3$\omega\rho$ models, with dashed lines indicating strong magnetic field impacts. The pressure shown is the effective pressure, $\tilde{P} = (3/2)P_\parallel - (1/2)P_\perp$, as a function of baryon number density $n_B$.}
    \label{fig:eos_comparison}
\end{figure}

We model the EoSs for the stellar crust and core regions, either separately or self-consistently. For all regions, leptons (electrons and muons) are modeled as a free Fermi gas under the influence of the magnetic field, as they do not interact with respect to the strong force. Nevertheless, nuclei, nucleons (baryon without strangeness, protons and neutrons) and in some cases also hyperons (baryons with net strangeness) are connected with leptons through the conditions of chemical equilibrium and charge neutrality. We do not account for temperature effects in the EoS in this work.

For the microscopic description, we assume the magnetic field $B$ to be (locally) pointing in the z-direction, inducing Landau quantization perpendicular to the magnetic field, in the x- and y- directions. This local ``z" direction does not directly correspond to any direction in particular within the star; it is merely an abstraction that allows us to calculate the thermodynamic quantities of stellar matter given local values of magnetic field and chemical potentials. We further include the effects of the anomalous magnetic moment (AMM) of the baryons and leptons, creating an asymmetry between the up and down spin states, that either align or anti-align with the local magnetic field, following Ref.~\citep{Strickland:2012vu} and references therein.

\subsection{NL3$\omega\rho$}

The NL3$\omega\rho$ model \citep{Horowitz:2000xj,Horowitz:2001ya,Pais:2016xiu} is a relativistic mean-field model of Walecka type with non-linear terms for the mesons, including a term that mixes the vector isoscalar  and the vector isovector fields ($\omega\rho$) to model the density dependence of the symmetry energy. It describes bulk matter made of nucleons interacting strongly through the scalar isoscalar meson $\sigma$ (mediating attraction), the vector isoscalar meson $\omega$ (mediating repulsion), and the vector isovector $\rho$ (for isospin asymetric matter), together with a free gas of electrons and muons. The model reproduces standard nuclear properties for isospin-symmetric matter at saturation: binding energy per baryon $B/A=-16.24$ MeV, saturation density $\rho_0=0.148$ fm$^{-3}$, normalized nucleon effective mass $M^*/M=0.60$, incompressibility $K=270$ MeV and symmetry energy $\mathcal{E}_{sym}=31.66$ MeV with slope $L=55$ MeV, in addition to being able to describe 2$M_\odot$ stars.

\subsubsection{Core}

Using the same construction as in \cite{Scurto:2022vqm,Scurto:2024xxo} the Lagrangian density of the NL3$\omega\rho$ model under the influence of magnetic fields is given by 
\begin{equation}
\mathcal{L}=\sum_{i=p,n}\mathcal{L}_i+\mathcal{L}_l+\mathcal{L}_\sigma+\mathcal{L}_\omega+\mathcal{L}_\rho+\mathcal{L}_{nl}\, .
\end{equation}
The nucleons and lepton terms are given by
\begin{align}
    \mathcal{L}_i&=\bar{\psi}_i\big[\gamma_\mu iD^\mu-M_*-\frac{1}{2}\kappa_i\sigma_{\mu\nu}F^{\mu\nu}\big]\psi_i \, \\
    \mathcal{L}_l&=\sum_{l=e,\mu}\bar{\psi}_l\big[\gamma_\mu\big(i\partial^\mu+eA^\mu\big)-m_l\big]\psi_l \, , 
\end{align}
with $i=p,n$ and $l=e, \mu$. $\kappa_i$ is the AMM coupling (see Tab.~\ref{kappatable}). $m_l$ is the lepton mass and
\begin{equation}
    M_*=M-g_\sigma\phi \, ,
\end{equation}
is the nucleon effective mass. The covariant derivative for the nucleons depends on the isoscalar and isovector vector interactions, in addition to the electromagnetic coupling
\begin{equation}
    iD^\mu=i\partial^\mu-g_\omega V^\mu-\frac{g_\rho}{2}\mathbf{\tau}\cdot\mathbf{b}^\mu-\frac{1+\tau_3}{2}eA^\mu \, ,
\end{equation}
where $e=\sqrt{4\pi/137}$ is the electron charge. The magnetic vector potential is $A^\mu=(0,0,Bx,0)$, i.e. $B$ is oriented along the $z$ axis.

The terms for the scalar, vector and isovector mesons are given by
\begin{align}
    \mathcal{L}_\sigma&=\frac{1}{2}\bigg(\partial_\mu\phi\partial^\mu\phi-m_\sigma^2\phi^2 \bigg) \, ,\\
    \mathcal{L}_\omega&=-\frac{1}{4}\Omega_{\mu\nu}\Omega_{\mu\nu}+\frac{1}{2}m_\omega^2V_\mu V^\mu  \, ,\\
    \mathcal{L}_\rho&=-\frac{1}{4}\mathbf{B}_{\mu\nu}\cdot\mathbf{B}^{\mu\nu}+\frac{1}{2}m_\rho^2\mathbf{b}_\mu\cdot \mathbf{b}^\mu \, ,
\end{align}
with the tensors defined as
\begin{align}
F_{\mu\nu}&=\partial_\mu A_\nu-\partial_\nu A_\mu\, , \\
\Omega_{\mu\nu}&=\partial_\mu V_\nu - \partial_\nu V_\mu \, , \\
\mathbf{B}_{\mu\nu}&=\partial_\mu \mathbf{b}_\nu - \partial_\nu \mathbf{b}_\mu - g_\rho \left(\mathbf{b}_\mu \times \mathbf{b}_\nu \right)\, .
\end{align}
The non-linear terms of the interaction read
\begin{align}
    \mathcal{L}_{int}=-\frac{1}{6}\kappa\phi^3
    -\frac{1}{24}\lambda\phi^4+\Lambda_{\omega\rho}g_\omega^2g_\rho^2V_\mu V^\mu\mathbf{b}_\mu\cdot \mathbf{b}^\mu \, .
\end{align}

The neutron density follows
\begin{align}
    \rho_n&=\frac{1}{2\pi^2}\sum_s\bigg[\frac{{k_{n,s}^{F}}^3}{3}-\frac{1}{2}s\kappa_nB\bigg(\bar{m}_nk_{n,s}^{F}\notag\\
    &+E_n^{F2}\bigg(\arcsin\bigg(\frac{\bar{m}_n}{E_n^F}\bigg)-\frac{\pi}{2}\bigg)\bigg)\bigg]\, ,
\end{align}
where the sum is over the spin states, $E_n^F$ is the Fermi energy and the Fermi momentum is given by
\begin{equation}
    k_n^{F2}=E_n^{F2}-(M_*-s\kappa_nB)^2\, .
    \label{Eq.fermi_mom}
\end{equation}
For the protons and leptons, that interact with the $B$-field, the density is given by
\begin{equation}
    \rho_i=\frac{|q|B}{2\pi^2}\sum_{\nu=0}^{\nu_{\rm max}^i}\sum_s k^F_{i,\nu s} \, ,
\end{equation}
with $i=p,l$, where $\nu=n+\frac{1}{2}-\frac{1}{2}\frac{q}{|q|}s=0,1,\cdots,\nu_{\rm max}$ enumerates the Landau levels (LLs) for fermions with electric charge $q$, $\nu_{\rm max}$ is the largest LL 
occupied by fully degenerate charged fermions, defined as
\begin{align}
    \nu_{\rm max}^i=\frac{(E_i^{F}+s\kappa_iB)^2-m^{2}}{2 |q| B} \, ,  
\end{align}
and the Fermi momentum for charged particles is given by
\begin{align}
    k^{F2}_{i,\nu s}=E_i^{F2}-(\sqrt{m^{2}- 2\nu |q| B}-s\kappa_iB)^2 \,,
    \label{Eq.fermi_mag}
\end{align}
where for the electrons the AMM is set to zero.
The chemical potentials for the protons and neutrons are then given respectively by
\begin{align}
    \mu_p=&E_p^F+g_\omega V^0+\frac{1}{2}g_\rho b^0 \, , \\
    \mu_n=&E_n^F+g_\omega V^0-\frac{1}{2}g_\rho b^0 \, .
\end{align}

The total energy density of the magnetized core EoS is given by
\begin{equation}    
\mathcal{E}=\mathcal{E}_{kin}^p+\mathcal{E}_{kin}^n+\mathcal{E}_{kin}^l+\mathcal{E}_F \, , 
 \label{Eq_energy1}
\end{equation}
with $\mathcal{E}_{F}$ the field contribution written as
\begin{align} 
    \mathcal{E}_{F}&=\frac{m_\omega^2}{2}\omega_0^2+\frac{m_\rho^2}{2}b_{3,0}^2+\frac{m_\sigma^2}{2}\phi_0^2 \nonumber \\
+&\frac{\kappa}{6}\phi_0^3+\frac{\lambda}{24}\phi_0^4+3\Lambda_{\omega\rho} g_\rho^2g_\omega^2\omega_0^2b_{3,0}^2 \, , \label{Eq_energy2}
\end{align}
and $\mathcal{E}_{kin}^j$ the single-particle energies of particles $j=p,n,l$
\begin{align} 
    \mathcal{E}_{kin}^n&= \frac{1}{4\pi^2}\sum_s\bigg[\frac{1}{3}k_{F}^nE_n^{F3}-\frac{2}{3}s\kappa_nBE_n^{F3}\bigg(\arcsin\bigg(\frac{\bar{m}_n}{E_n^F}\bigg)\notag\\&-\frac{\pi}{2}\bigg)-\bigg(\frac{1}{3}s\kappa_nB+\frac{1}{4}\bar{m}_n\bigg)\bigg(\bar{m}_nk_{F}^nE_n^{F}\notag\\&+\bar{m}_n^3\ln\bigg|\frac{k_{F}^n+E^F_n}{\bar{m}_n}\bigg|\bigg)\bigg], \label{Eq_energy3}
\end{align}
\begin{align} 
    \mathcal{E}_{kin}^i&=\frac{|q|B}{4\pi^2}\sum_{\nu=0}^{{\nu_{\rm max}^i}}\sum_s\bigg[k_{F,\nu s}^iE_i^F+\bigg(\sqrt{m^{2}+2\nu |q|B}\notag\\ &-s\kappa_iB\bigg)^2 \ln\bigg|\frac{k_{F,\nu s}^i+E^F_i}{\sqrt{m^{2}+2\nu |q|B}-s\kappa_iB}\bigg|\bigg] \, ,  \label{Eq_energy4}
\end{align}
with $i=p,l$, where again the AMM of leptons is neglected.
The pressure is obtained from the thermodynamic relation
\begin{equation}
   P=\mu_p\rho_p+\mu_n\rho_n+\mu_e(\rho_e+\rho_\mu)-\mathcal{E}. \label{Ppar}
\end{equation}

Since the magnetic field induces anisotropy in the energy-momentum tensor, one must calculate the perpendicular component of the pressure. The thermodynamic pressure, Eq.~(\ref{Ppar}), is equal to the parallel component. 
The perpendicular component reads:
\begin{equation}
   P_{\perp}=P-\mathcal{M}B \, , 
   \label{pperpH}
\end{equation}
with the magnetization $\mathcal{M}$ given by
\begin{equation}
 \mathcal{M}=-\frac{\partial \mathcal{E}}{\partial B} \, .    
\end{equation}

\subsubsection{Inner crust}
In the inner crust, due to the competition between the strong and Coulomb forces, geometrical structures of nucleons are formed, immersed in a sea of neutrons and electrons. We calculate this region of the star from a compressible liquid drop model (CLDM) approximation, where the Gibbs equilibrium conditions are imposed at the intersection between the dense (clusters) and gas (free nucleons). The minimization of the total energy density takes into account the Coulomb and surface terms. %

The total energy density of the system in the inner crust is given by:
\begin{equation}
\mathcal{E}=f\mathcal{E}^I+(1-f)\mathcal{E}^{II}+\mathcal{E}_{Coul}+\mathcal{E}_{surf}+\mathcal{E}_e \,, \label{enerCrust}
\end{equation}
where the $\mathcal{E}$ are the bulk energy density of protons and neutrons in the dense ($I$) and gas ($II$) phases, {calculated using the magnetized NL3$\omega\rho$ model following Eq.~\ref{Eq_energy1}}, $f$ is the fraction of the heavy cluster in the dense phase, $\mathcal{E}_e$ is the energy density of the electrons, and $\mathcal{E}_{Coul}$ and $\mathcal{E}_{surf}$ are the Coulomb and surface energy density terms, respectively, given by
\begin{align}
\mathcal{E}_{Coul}&=2\alpha e^2\pi\Phi R_d^2 \left(\rho_p^I-\rho_p^{II}\right)^2 \,,  \\
\mathcal{E}_{surf}&=\frac{\sigma\alpha D}{R_d} \,,
\end{align}
with
\begin{align}
\Phi&=\left(\frac{2-D\alpha^{1-2/D}}{D-2}+\alpha\right)\frac{1}{D+2} \, , \qquad D=1,3 \, ,  \nonumber \\
\Phi&=\frac{\alpha-1-\ln\alpha}{D+2} \, , \qquad  D=2 \, ,
\end{align}
where $D$ is the dimension of the geometry, $\alpha=f$ for droplets, rods and slabs, and $\alpha=1-f$ for bubbles and tubes, and $\sigma$ is the surface tension functional, that is calculated from a fit to a Thomas-Fermi calculation \citep{Avancini_2012}.  
From the minimization of Eq.~(\ref{enerCrust}), we obtain the following relation {between the surface and the Coulomb corrections, the expression for the radius of the cluster}%
\begin{align}
&\mathcal{E}_{surf}=2\mathcal{E}_{Coul} \, ,   \\
&R_d=\left[\frac{\sigma D}{4\pi e^2\Phi\left(\rho_p^I-\rho_p^{II}\right)^2 }\right]^{1/3} \, ,
\end{align}
{and the Gibbs equilibrium, given by}
\begin{align}
    \mu_n^I&=\mu_n^{II} \, , \\
    \mu_p^I&=\mu_p^{II}-\frac{\mathcal{E}_{surf}}{(1-f)f(\rho^I_p-\rho^{II}_p)} \, , \\
    P^I&=P^{II}\\&+\mathcal{E}_{surf}\bigg[\frac{3}{2\alpha}\frac{\partial\alpha}{\partial f}+\frac{1}{2\Phi}\frac{\partial\Phi}{\partial f}-\frac{((1-f)\rho_p^I+f\rho_p^{II})}{(1-f)f(\rho^I_p-\rho^{II}_p)}\bigg] \, .
\end{align}
For further details, the reader can consult \citet{Scurto:2022vqm}. As mentioned in Sec.~\ref{sec:eos}, this description for the inner crust of neutron stars is combined with the SLy4 EoS for the outer crust \citep{Douchin:2001sv}.

\subsection{CMF} %

To describe bulk matter interacting with respect to the strong force, but more realistically describing large density, we make use of the Chiral Mean Field (CMF) model. It is a relativistic microscopic model that describes nucleons, hyperons, and quarks interacting strongly through scalar mesons (mediating attraction) and vector mesons (mediating repulsion), together with a free gas of electrons and muons \citep{Dexheimer:2008ax}. Most importantly, the model reproduces the chiral symmetry restoration phase transition and the deconfinement phase transition to quark matter \cite{Dexheimer:2009hi}, two important feature of Quantum Chromodynamics (QCD), in the high-energy regime. In this work, we do not include finite-temperature effects in the EoS, so high energy means high baryon chemical potential $\mu_B$, related (in a complex way that depends on the EoS) to the baryon number density $n_B$.

The CMF model is based on a nonlinear realization of the chiral linear sigma model \citep{Papazoglou:1997uw}. Like other relativistic mean-field (RMF) models, it uses mean-field mesons to approximate the effects of gluons, simplifying considerably the calculations of QCD. On the other hand, chiral models differ from RMF Walecka-type models, as they have particle masses generated by the medium. In this work, we do not include quarks in order to disentangle the effects of magnetic field and deconfinement. See Ref.~\citep{Peterson:2023bmr} for a discussion on how magnetic fields and temperature affect the dense-matter EoS described by chiral models with quarks. 
The model reproduces standard nuclear properties for isospin-symmetric matter at saturation: binding energy per baryon $B/A=-16$ MeV, saturation density $\rho_0=0.15$ fm$^{-3}$, incompressibility $K=300$ MeV and symmetry energy $\mathcal{E}_{sym}=30$ MeV with slope $L=88$ MeV, 
in addition to being able to describe 2$M_\odot$ stars (see Ref.~\cite{Dexheimer:2018dhb} for a version of CMF with $\omega\rho$ interaction, but no magnetic field effects).

The CMF Lagrangian density, following the CMF formalism with magnetic-field effects presented in \citep{Dexheimer:2011pz} is
\begin{equation}
\mathcal{L}=\mathcal{L}_{kin}+\mathcal{L}_{int}+\mathcal{L}_{scal}+\mathcal{L}_{vec}+\mathcal{L}_{\rm esb}\,,
\end{equation}
where the kinetic energy term for baryons is
\begin{equation}
\mathcal{L}_{kin}=\sum\limits_{i\in B}\left(\bar{\psi}_i i\left(\gamma_\mu\partial^\mu +\frac{1}{2} \kappa_i\sigma^{\mu\nu}F_{\mu\nu}\right)\psi_i\right)\,,
\end{equation}
the baryon-meson interaction term is 
\begin{align}\label{massCMF}
\mathcal{L}_{int}=-\sum\limits_{i\in B}\bigg(\bar{\psi}_i[\gamma_0(g_{i\omega}\omega+g_{i\rho}\rho+g_{i\phi}\phi)+m_i^*]\psi_i\bigg)\,,\nonumber\\
\end{align}
the scalar meson self-interaction term is
\begin{align}
\mathcal{L}_{scal}&=-\frac{1}{2}k_0\chi_0^2(\sigma^2+\delta^2+\zeta^2)+k_1(\sigma^2+\delta^2+\zeta^2)^2\nonumber\\&+k_2\bigg(\frac{\sigma^4+\delta^4}{2}+\zeta^4+3(\sigma\delta)^2\bigg)+k_3\chi_0(\sigma^2-\delta^2)\zeta\nonumber\\&-k_4\chi_0^4
+\frac{\epsilon}{3}\chi_0^4\ln\bigg(\frac{(\sigma^2-\delta^2)\zeta}{\sigma_0^2\zeta_0}\bigg)\,,
\end{align}
the vector meson self-interaction term is
\begin{align}
\mathcal{L}_{vec}&=\frac{1}{2}(m_\omega^2\omega^2+m_\rho^2\rho^2+m_\phi^2\phi^2)\nonumber\\
&+g_4\left(\omega^4+\frac{\phi^4}{4}+3\omega^2\phi^2+\frac{4\omega^3\phi}{\sqrt{2}}+\frac{2\omega\phi^3}{\sqrt{2}}\right)\,,
\end{align}
and the term corresponding to an explicit breaking of chiral symmetry is
\begin{equation}
\mathcal{L}_{\rm esb}=-\bigg(m_{\pi}^{2}f_{\pi}\sigma+(\sqrt{2}m_{k}^{2}f_{k}-\frac{1}{\sqrt{2}}m_{\pi}^{2}f_{\pi})\zeta\bigg)\,\,,
\end{equation}
in addition to a constant term. In these expressions $\psi_i$ is the baryon or lepton wave function, $\gamma_\mu$ represents the Dirac matrices, $\sigma$, $\delta$, and $\zeta$ are scalar mesons, while $\omega$, $\rho$, and $\phi$ are the vector mesons; $g_{ij}$ are coupling constants between baryons $i$ and mesons $j$ (see Tables in Ref.~\cite{Cruz-Camacho:2024odu} for coupling values and additional details). 
Furthermore, $m_i^*$ is the effective mass of baryon $i$, $\sigma^{\mu\nu}=\frac{i}{2}[\gamma^\mu,\gamma^\nu]$, and $F^{\mu\nu}$ is the electromagnetic tensor. The AMM couplings $\kappa_i$ are shown in Tab.~\ref{kappatable}.

\begin{table*}[t]
\caption{Table of AMM coupling strengths for leptons and baryons~\citep{Zyla:2020zbs}. $\kappa_i$ is equal to the product of the AMM coupling $\mathcal{K}_i$ and an appropriate magneton. For baryons, this is the nuclear magneton $\frac{e}{2m_p}$, whereas, for leptons the magneton is $\frac{e}{2m_i}$, yielding unique values for the electron and muon. In the magneton expressions, $e$ is the elementary charge and the masses are the vacuum masses.}
\def\arraystretch{1.8}
\begin{tabular}{ccccccccccc}
\hline
\hline
i & e & $\mu$ & p & n &  $\Lambda$ \ &\ \ $\Sigma^+$ \ \ &\ \ $\Sigma^0$ \ \ & \ \ $\Sigma^-$ \ \ &\ \ $\Xi^0$ \ \ &\ \ $\Xi^-$\\
$\mathcal{K}_i$ & $0.00116$ \ \ &\ \ $0.001166$ & $1.79$ &\ $-1.91$ &  $-0.61$ & $1.67$ & $1.61$ & $-0.38$ & $-1.25$ & $0.06$\\
\hline
\hline
\end{tabular}
\label{kappatable}
\end{table*}

The Lagrangian density cannot be fully decoupled between baryons and mesons. Instead, the result is a relativistic Fermi gas with an effective mass generated by the scalar meson fields $m_i^*=g_{i\sigma}\sigma+g_{i\delta}\delta+g_{i\zeta}\zeta+\Delta m_{i}$\footnote{$\Delta m_{i}$ is the bare mass, which does not come from the mean fields. It has values $150$ MeV for nucleons and $342.3$ MeV for hyperons.} and effective Fermi energy modified by the vector meson fields $E_i^*=E_i-g_{i\omega}\omega-g_{i\rho}\rho-g_{i\phi}\phi$. At $T=0$, the Fermi energy of particles corresponds to their chemical potentials $E_i\rightarrow\mu_i=B_i\mu_B+Q_i\mu_Q$, where $B_i$, is the particle baryon number ($1$ for baryons and $0$ for leptons), $Q_i$ the particle electric charge, $\mu_B$ the baryon chemical potential, and $\mu_Q$ the charge chemical potential. While $\mu_B$ is our independent variable, $\mu_Q$ is determined by imposing charge neutrality and chemical equilibrium with leptons. The mass term is obtained for leptons from Eq.~\ref{massCMF} setting $g_i=0$, $m*_i=m_i$, in addition to $\mu^*_i=\mu_i$.

Additionally, $\bar{m}$ is the particle effective mass modified by the magnetic field given by $\bar{m_i}=\sqrt{m_i^{*2}+2\nu|q_i|B}-s\kappa_i B$ for charged particles and $\bar{m_i}=m_i^{*}-s\kappa_i B$ for uncharged particles, $\nu=n+\frac{1}{2}-\frac{s}{2}\frac{q_i}{|q_i|}$ is the Landau level, with n being the discretized orbital angular momentum of the particle in the transverse plane, and $s=\pm1$ the spin projection of the particle along the direction of the magnetic field.

From here, we obtain the expressions for number density, scalar density, energy density, parallel pressure (z-direction), and perpendicular pressure (x- and y- directions) of charged particles~\citep{Strickland:2012vu}:
\begin{align}
    n_i&=\frac{|q_i|B}{2\pi^2}\sum_{s=\pm1}\sum_{\nu\leq\nu_{max}}k_{z_{i}}\,,\\
    n_{S,i}&=\frac{|q_i|B}{2\pi^2}\frac{\bar{m_i}m_i^*}{\bar{m_i}+s_i\kappa_i B}\ln\left(\frac{\sqrt{k_{z_{i}}^2+\mu^*_i}}{\bar{m_i}}\right)\,,
\end{align}
\begin{align}
    \varepsilon_i&=\frac{|q_i|B}{4\pi^2}\sum_{s=\pm1}\sum_{\nu\leq\nu_{max}}\left[\mu^*_i k_{z_{i}}+\bar{m_i}^2\ln\left(\frac{\mu^*_i+k_{z_{i}}}{\bar{m_i}}\right)\right]\,,
    \\
    P_{||i}&=\frac{|q_i|B}{4\pi^2}\sum_{s=\pm1}\sum_{\nu\leq\nu_{max}}\left[\mu^*_i k_{z_{i}}-\bar{m_i}^2\ln\left(\frac{\mu^*_i+k_{z_{i}}}{\bar{m_i}}\right)\right]\,,
    \\
    P_{\perp i}&=\frac{|q_i|B^2}{2\pi^2}\sum_{s=\pm1}\sum_{\nu\leq\nu_{max}}\Bigg[\Bigg(\frac{|q_i|\nu\bar{m_i}}{\sqrt{m_i^{*2}+2\nu|q_i|B}}\nonumber\\&-s\kappa_i\bar{m_i}\Bigg)
    \ln\left(\frac{\mu^*_i+k_{z_{i}}}{\bar{m_i}}\right)\Bigg]\,,
    \label{3}
\end{align}
where the Fermi momentum of particle $i$ in the local magnetic-field direction is $k_{z_{i}}=\sqrt{{\mu^*_i}^2-\bar{m_i}^2}$ and $\nu_{max}=\left\lfloor\frac{(\mu^*_i+s\kappa_i B)^2-m_i^{*2}}{2|q_i|B}\right\rfloor$ is the largest Landau level for which $k_{z_{i}}$ is real. For uncharged particles, these equations are instead
\begin{align}
    n_i&=\frac{1}{2\pi^2}\sum_{s=\pm1}\Bigg[\frac{k_i^3}{3}\nonumber\\&-\frac{s\kappa_i B}{2}\Bigg(m_i^* k_i{\mu^*}^2\Bigg(\arcsin\left(\frac{m_i^*}{\mu^*_i}\right)-\frac{\pi}{2}\Bigg)\Bigg)\Bigg]\,,\\
    n_{S,i}&=\frac{1}{2\pi^2}m_i^*\Bigg(\frac{k_i\mu^*_i}{2}-\frac{(m_i^{*}-s\kappa_i B)^2}{2}\nonumber\\
    &\times\ln\left(\frac{k_i+\mu^*_i}{m_i^*-s\kappa_i B}\right)\Bigg)\,,\\
    \varepsilon_i&=\frac{1}{48\pi^2}\sum_{s=\pm1}\bigg[\mu^*_i k_i\left(6{\mu^*_i}^2-3\bar{m_i}^2-4s\kappa_i B\bar{m_i}\right)\nonumber\\&-8s\kappa_i B{\mu^*_i}^3\left(\arcsin\left(\frac{m_i^*}{\mu^*_i}\right)-\frac{\pi}{2}\right)\nonumber\\&-\bar{m_i}^3(3\bar{m_i}+4s\kappa_i B)\ln\left(\frac{\mu^*_i+k_i}{\bar{m_i}}\right)\bigg]\,,\nonumber
\end{align}
\begin{align}
    P_{||i}&=\frac{1}{48\pi^2}\sum_{s=\pm1}\bigg[\mu^*_i k_i\left(2{\mu^*_i}^2-5\bar{m_i}^2-8s\kappa_i B\bar{m_i}\right)\nonumber\\&-4s\kappa_i B{\mu^*_i}^3\left(\arcsin\left(\frac{m_i^*}{\mu^*_i}\right)-\frac{\pi}{2}\right)\nonumber\\&+\bar{m_i}^3(3\bar{m_i}+4s\kappa_i B)\ln\left(\frac{\mu^*_i+k_i}{\bar{m_i}}\right)\bigg]\,,\\
    P_{\perp i}&=\frac{1}{48\pi^2}\sum_{s=\pm1}\bigg[\mu^* k_F(2{\mu^*}^2-5\bar{m}^2-12s\kappa B\bar{m}\nonumber\\&-12(s\kappa B)^2)+3\bar{m}^2(\bar{m}+2s\kappa B)^2\ln\left(\frac{\mu^*+k_F}{\bar{m}}\right)\bigg]\,.
\end{align}
where the parallel and perpendicular pressures specify the components of the energy-momentum tensor of matter in the local rest frame of the system.
The total baryon number density is then the sum over all particle number densities multiplied by their baryon number $B_i$, $n_B=\sum_i B_in_i$. In addition to a sum over particles, matter energy density and pressures receive additional terms from the mesons (with the vacuum subtracted) and free leptons. 
At $T=0$, mesons provide no contribution to the kinetic term, so their contributions are simply $\mathcal{L}_{\rm mesons}=\mathcal{L}_{\rm scal}+\mathcal{L}_{\rm vec}+\mathcal{L}_{\rm esb} -\mathcal{L}_{\rm vacuum}$. The total energy density, parallel pressure, and perpendicular pressure then become
\begin{align}
&\varepsilon=\sum\limits_{i\in B,\rm{lep}}\varepsilon_i+\varepsilon_{\rm int}-\mathcal{L}_{\rm mesons}\,, \label{eq:Fermisums1}\\
&P_{||}=\sum\limits_{i\in B,\rm{lep}}P_i+\mathcal{L}_{\rm mesons}\,, \label{eq:Fermisums2}\\
&P_{\perp}=\sum\limits_{i\in B,\rm{lep}}P_i+\mathcal{L}_{\rm mesons},
\,,
\label{eq:Fermisums3}
\end{align}
with $\varepsilon_{\rm int} = \sum\limits_{i \in B} \left( g_{i \omega} \omega +  g_{i \phi} \phi  + g_{i\rho} \rho  \right) n_{i}$.

{In Fig. \ref{fig:eos_comparison}  we compare these two EoS, which differ in the core and inner crust region. The outer core was smoothly transitioned to SLy4 \citep{Douchin:2001sv}.
We show the thermodynamic pressure as function of the density for a value of the magnetic field of $10^{16}$G. Even though NL3$\omega\rho$ has a smaller $L$ at saturation density, it becomes stiffer than CMF because this one includes exotic degrees of freedom. The CMF presents a first-order phase transition at very low densities, reminiscent of the nuclear liquid-gas phase transition for isospin symmetric matter.

\section{Primitive inversion scheme}
\label{app:inversion}

One conceptual challenge arises in the use of standard primitive inversion
algorithms, since the EoS now acquires a velocity dependence
through the use of the comoving magnetic field strength, $b$. In other words, the equations acquire a fluid frame dependence that needs to be accounted for.
Rather than modifying the algorithm~\citep{Kastaun:2020uxr} itself, 
we make use of the fact that the anisotropic pressure corrections
are intrinsically small ($<10\%$).
While this overall leads to a significant increase in computational cost, it is sufficient for a first exploration of this effect.

We call $Y_B=\sqrt{b_\mu b^\mu}$ the effective magnetic field composition variable, such that the EoS we use is given by
\begin{align}
    \tilde{P}= \tilde{P}_{B}\left(\rho, Y_B\right) + \rho \varepsilon \left(\Gamma_{\rm th} -1\right)\,,
\end{align}
where the second term has been added to approximately capture thermal effects \citep{Bauswein:2010dn}. We nominally choose $\Gamma_{\rm th}=1.8$.

Our modified scheme operates as follows:
\begin{enumerate}
    \item Choose an initial guess $Y_B = B$, and set the magnetic field $\bar{B}^i = B^i$ to be used in the next step. 
    \item Solve the primitive inversion \citep{Kastaun:2020uxr} for fixed $Y_B$, and obtain an approximate value, $\bar{v}^i$, for the velocity.
    \item Recompute $\bar{Y}_B = \sqrt{\bar{b}^2}$, where $\bar{b}^2 = \frac{B^2}{\bar{W}^2} + \left(B^i \bar{v}_i\right)^2$ is the comoving magnetic field strength, and $\bar{W}^{-1} = \sqrt{1-\bar{v}_i \bar{v}^i}$ is the inverse Lorentz factor, both computed from $\bar{v}^i$.
    \item Using the approximate compositional value of $\bar{Y}_B$, compute the polarization $\bar{\kappa}$, and use it to rescale the magnetic field $\bar{B}^i = \sqrt{1-\kappa}\, B^i$.
    \item Iterate steps 2.-4. until convergence in $\bar{Y}_B$, which empirically takes about five iterations in regions of high magnetization.
\end{enumerate}

\bibliography{main}{}
\bibliographystyle{aasjournalv7}

\end{document}